\documentclass[12pt,english]{iopart}
\usepackage[T1]{fontenc}
\usepackage[utf8x]{inputenc}
\usepackage[english]{babel}
\usepackage{cite}
\usepackage{graphicx}
\usepackage{amssymb}
\usepackage{setstack}
 \usepackage{amsfonts}
 \usepackage{dsfont}
 \usepackage{epsfig}
 \usepackage{subfigure}
 \usepackage{graphicx}
 \usepackage{float}

\makeatletter
\@ifundefined{textcolor}{}
{%
 \definecolor{BLACK}{gray}{0}
 \definecolor{WHITE}{gray}{1}
 \definecolor{RED}{rgb}{1,0,0}
 \definecolor{GREEN}{rgb}{0,1,0}
 \definecolor{BLUE}{rgb}{0,0,1}
 \definecolor{CYAN}{cmyk}{1,0,0,0}
 \definecolor{MAGENTA}{cmyk}{0,1,0,0}
 \definecolor{YELLOW}{cmyk}{0,0,1,0}
 }
\usepackage{color} 
\renewcommand{\vec}[1]{\mathbf{#1}}
\renewcommand{\text}[1]{{\rm{#1}}}

\newcommand{\ket}[1]{\left|#1\right>}
\newcommand{\bra}[1]{\left<#1\right|}

\newcommand{\sumint}[2]
{{\textstyle \sum}\hspace{-1.1em}{\displaystyle\int \limits_{#1}^#2}}

\makeatother

\usepackage{babel}

\begin{document}

\title[A comparative analysis of binding in ultralong-range Rydberg molecules]
{A comparative analysis of binding in ultralong-range Rydberg molecules}

\author{C Fey$^1$, M Kurz$^1$, P Schmelcher$^{1,2}$, S T Rittenhouse$^3$ and H R Sadeghpour$^4$ }
\address{$^1$Zentrum f\"{u}r Optische Quantentechnologien, Universit\"{a}t
Hamburg, Luruper Chaussee 149, 22761 Hamburg, Germany}
\address{$^2$The Hamburg Centre for Ultrafast Imaging,
Luruper Chaussee 149,
22761 Hamburg, Germany}
\address{$^3$Department of Physics and Astronomy, Western Washington University, 516 High Street, Bellingham, WA 98225, USA}
\address{$^4$ITAMP, Harvard-Smithsonian Center for Astrophysics, 60 Garden Street, Cambridge, MA 02138, USA}
\date{\today}
\begin{abstract}
We perform a comparative analysis of different computational approaches employed to explore the electronic structure of ultralong-range Rydberg molecules. Employing the Fermi pseudopotential approach, where the interaction is approximated by an $s$-wave bare delta function potential, one encounters a non-convergent behavior in basis set diagonalization. Nevertheless, the energy shifts within the first order perturbation theory coincide with those obtained by an alternative approach relying on Green's function calculation with the quantum defect theory. A pseudopotential that yields exactly the results obtained with the quantum defect theory, i.e. beyond first order perturbation theory, is the regularized delta function potential. The origin of the discrepancies between the different approaches is analytically motivated.
\end{abstract}

%\pacs{03.75.Lm, 67.85.De}

\maketitle

\section{Introduction}

Diatomic ultralong-range Rydberg molecules consisting of a Rydberg atom whose electron, upon frequent scattering off a ground state atom, binds the atom, localized at large distances ($\sim10^3$ {Bohr radii}), to the ion core of the Rydberg atom were predicted theoretically in Ref. \cite{greene_creation_2000}.  A subset of these Rydberg molecules with s-wave dominated electronic orbitals were later observed \cite{bendkowsky_observation_2009}.  This initial experimental work was followed by a flurry of new observations of Rydberg molecules with a variety of electronic and rovibrational structures \cite{bendkowsky_rydberg_2010,butscher_atommolecule_2010,
li_homonuclear_2011,tallant_observation_2012,
bellos_excitation_2013,
anderson_photoassociation_2014,krupp_alignment_2014,
sasmannshausen_experimental_2014}.
For electronic $s$-wave scattering of the  Rydberg electron from the ground state atom, two types of ultralong-range molecular states can be distinguished: the non-polar, low angular momentum quantum defect states and the polar, high angular momentum ``trilobite'' states. Theoretically, some aspects of these high angular momentum states have been explored, notably the semiclassical nature of the rich oscillatory structure in the adiabatic potential energy surfaces \cite{granger_quantum_2001}, their large permanent electric dipole moments \cite{sadeghpour_how_2013}, as well as the precise control over their electronic properties and molecular orientation by static electric and magnetic fields \cite{kurz_ultralong-range_2014,kurz_electrically_2013}. 

Further emphasis has been laid on polyatomic systems consisting of a Rydberg atom bound to two or more ground state atoms \cite{liu_polyatomic_2006,liu_ultra-long-range_2009} or to a diatomic polar perturber \cite{rittenhouse_ultracold_2010,rittenhouse_ultralong-range_2011}.
Theoretical approaches to describe ultralong-range Rydberg molecules can be divided into two categories: methods using the Fermi pseudopotential\cite{greene_creation_2000,hamilton_shape-resonance-induced_2002} and methods using the  quantum defect approach \cite{khuskivadze_adiabatic_2002}- for a recent overview, we refer the reader to \cite{marcassa_interactions_2014}.
Although the two approaches are expected to be equivalent, they indeed differ in numerical implementation \cite{hamilton_shape-resonance-induced_2002} and are even used alternatively \cite{bendkowsky_rydberg_2010}. 

In this work we therefore aim at a concise analysis of the origin of the discrepancies obtained from these two approaches. We thereby point out the interconnection and limitations in both approaches which should be taken into consideration in future studies of ultralong-range Rydberg molecules. Specifically, we show that a pseudopotential modeled by a bare delta function potential leads to non-converging molecular potential energy curves in basis set diagonalization of the electronic Hamiltonian. Modeling the pseudopotential instead by a regularized delta function reproduces potential energy curves which agree exactly with those obtained by the quantum defect theory approach. Throughout our analysis we not only give limits of the bare delta function potential but also its validity as an approximation in first order perturbation theory. This ultimately links the two approaches.
%%%%%%%%%%%%%%%%%%%%%%%%%%%%%%%%%%%%%%%%%%%%%%%%%%%%%%%%%%%%%%%%%%%%%%%%%%%%%%%%%%%%%%
\section{Ultralong-range Rydberg molecules}
%Let us briefly present the physical system and the basic assumptions made in both, the Fermi pseudopotential approach and the quantum defect approach. Atomic units are used throughout except where stated otherwise. 
We consider a Rydberg atom whose ionic core is located at the origin and a neutral ground state atom located at the position $\vec{R}$ within the Rydberg electron orbit. In the Born-Oppenheimer (BO) approximation, the electronic Hamiltonian for this Rydberg molecule reads
\begin{equation}
\hat{H}= \hat{H}^{0}+ V(\vec{R})
\label{eqn:Hamiltonian}
\end{equation}
where $\hat{H}^{0}$ is the Hamiltonian describing the Rydberg electron in its ionic core potential and $V(\vec{R})$ is the interaction between the Rydberg electron and the neutral ground state atom which we call from now on the perturber. If $V$ is  short-ranged, the corresponding electronic wave function $\Psi(\vec{r})$ outside the range of $V$ can be determined from the scattering phase shifts induced by $V$. Although a more general treatment is possible, here we focus exclusively on pure $s$-wave {electron-perturber} scattering. 

%On the one hand the phase shift appears in the stationary electronic wavefunction $\Psi(\vec{r})$ of energy $E$, that behaves in regions near $\vec{R}$ but outside the range of the perturbing potential $V$ as
In the region where the electron is near the perturber, but outside of the range of $V$, the electron wave function behaves as
\begin{equation}
\Psi(\vec{r})\propto \frac{\sin\left(\rho-\delta(k)\right)}{k\rho} ,
\label{eqn:psi_semiclassic}
\end{equation}
where $\rho= |\vec{r} -\vec{R}|$ is the relative electron-perturber coordinate and $\delta(k)$ is the $s$-wave phase shift depending on the wave vector $k$ given semiclassically as
\begin{equation}
k=k(R,E) = \sqrt{2 \left(E + \frac{1}{R}\right)} \ .
\label{eqn:k}
\end{equation}
The $s$-wave scattering length of the potential $V$ is obtained from $a[k]=-\tan( \delta(k))/k$, which in the low energy limit behaves as 
\begin{equation}
a[k]= a[0] + \frac{\pi}{3} \alpha[0] k + \mathcal{O}(k^2)
\label{eqn:scatlength}
\end{equation}
where $a[0]$ is the zero-energy scattering length and $\alpha[0]$ is the zero-frequency (static) polarizability of the ground state atom (polarized in the charge-atom interaction) \cite{bendkowsky_observation_2009,omont_theory_1977}. The momentum $k$ can be considered to be small (low-energy limit) because $R$ in (\ref{eqn:k}) is typically large and close to the classical turning points. The idea behind the Fermi pseudopotential approach is to replace $V$ by a zero-range pseudopotential that possesses the same $s$-wave scattering length as the original potential, while the quantum defect approach replaces the potential $V$ by the Dirichlet boundary condition (\ref{eqn:psi_semiclassic}).
\section{Bare delta function potential}
The pseudopotential usually employed to describe the $s$-wave binding of ultralong-range Rydberg molecules \cite{greene_creation_2000} is the bare delta function potential  
\begin{equation}
V(\vec{r},\vec{R})= 2 \pi a[k] \delta(\vec{r}-\vec{R}) ,
\label{eqn:pseudopotential}
\end{equation}
where the scattering length $a[k]$ depends on the energy $E$ and the perturber position $\vec{R}$ via (\ref{eqn:k}) and (\ref{eqn:scatlength}). In the case that one is interested only in energies $E$ close to some reference energy $E_i^0$ one may approximate
\begin{equation}
k \approx k(R,E_i^0) \ .
\label{eqn:k_fix} 
\end{equation}
We now consider the energy shift of the electronic Rydberg orbit by the perturbing potential $V$ where $\ket{\varphi_i}$ is an eigenstate of the bare Rydberg Hamiltonian with eigen energy $E_i^0$:
\begin{equation}
\hat{H}^0 \ket{\varphi_i}= E_i^0 \ket{\varphi_i} \ .
\end{equation}
For a compact notation we choose the basis states $\ket{\varphi_i}$ such that they diagonalize $V$ inside energetically degenerate submanifolds, i.e. $\bra{\varphi_i}V \ket{\varphi_j} \propto \delta_{ij}$ if $E^0_i=E^0_j$.
Using (\ref{eqn:k_fix}), we deduce from (\ref{eqn:pseudopotential}) the first order perturbation theory correction to $E_i^0$ as
\begin{equation}
E_i^{(1)}(\vec{R})=  2 \pi a\left[k(R,E_i^0)\right] \left|\varphi_i(\vec{R})\right|^2 ,
\label{eqn:first_order_fermi}
\end{equation}
which is the BO potential, valid for low angular momentum quantum defect, as well as for high angular momentum ``trilobite'' states. 

As an example we consider the $Rb(35s)+Rb(5s)$ state studied in \cite{bendkowsky_observation_2009,bendkowsky_rydberg_2010}. The correspondent eigenstates of $\hat{H}^0$ are the quantum defect states $\ket{n l m}$ that possess the energies $E^0_{nlm}=-1/2(n-\Delta_l)^2$, where $\Delta_l$ denotes the $l$-dependent quantum defect. We use $\Delta_0=3.13$, $\Delta_1=2.65$, $\Delta_2=1.35$ and neglect the quantum defects of higher, i.e. $\Delta_{l>2}=0$, angular momentum states. Here $n$, $l$ and $m$ are the usual hydrogenic quantum numbers. The properties of the perturber $Rb(5s)$ enter in (\ref{eqn:scatlength}) via the dominating triplet $s$-wave scattering parameters
\begin{equation}
 a_{Rb}[0]=-16.1 \ a_0 \qquad \text{and}  \qquad  \alpha_{Rb}[0]=319.2 \ a_0^3 \ ,
 \label{eqn:parameters}
 \end{equation}
 where $a_0$ is the Bohr radius. The e$^-$-Rb(5s) triplet scattering length has been calculated in \cite{bahrim_3se_2001} and the polarizability is obtained from \cite{marinescu_dispersion_1994}.
 \begin{figure}[h]
\includegraphics[width= 0.9\linewidth]{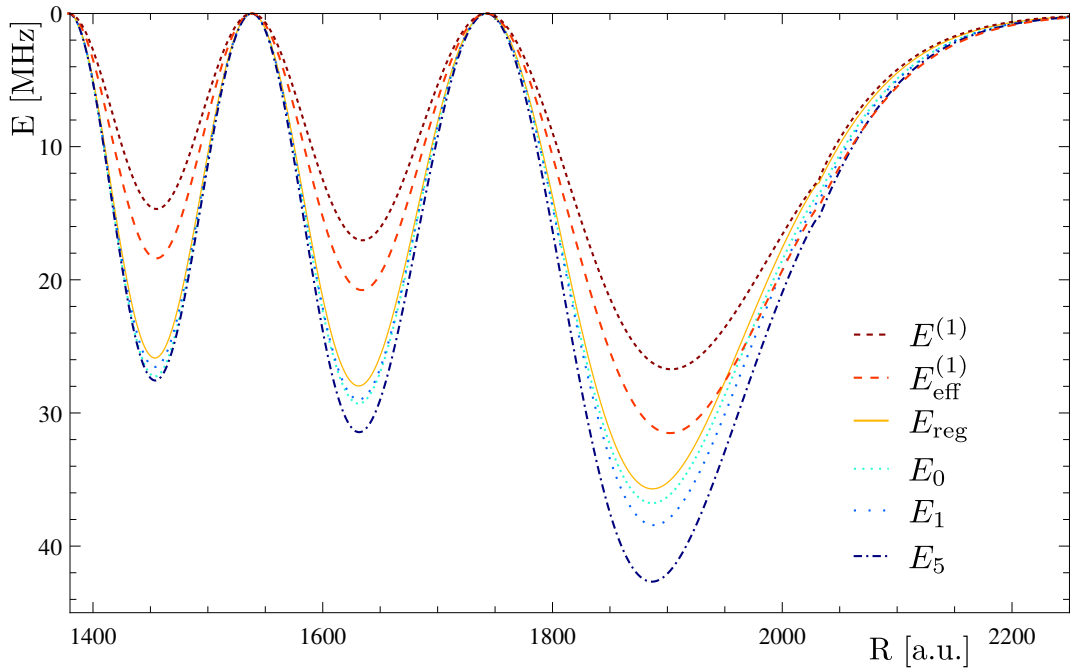}
\caption{Comparison of the Born-Oppenheimer potential curves $E$ as a function of the nuclear separation $R$ for a $Rb(35s)+Rb(5s)$ molecule with the parameters (\ref{eqn:parameters}) for different methods. The dissociation threshold is set to zero. $E^{(1)}$ and $E_\text{eff}^{(1)}$ are the first order approximations based on (\ref{eqn:first_order_fermi}), where $E_\text{eff}^{(1)}$ uses the effective zero-energy scattering length $a_\text{eff}[0]=-18.5 a_0$. The curve $E_\text{reg}(R)$ gives the energy for the regularized delta function potential based on (\ref{eqn:energy_Green}) while the three curves $E_{j}$ are determined via numerical diagonalization of (\ref{eqn:matrix}) including basis sets $32-j \leq n^* \leq 32+j$.}
\label{fig:comparison}

\end{figure}

The BO potential energy curve $E^{(1)}$ is shown (red dashed line) in Fig. \ref{fig:comparison}. At $R\approx 1900 a_0$,  it possesses a minimum that lies 26 MHz below the dissociation limit (here set to zero) and supports several bound states. The non-differentiability around $R\approx 2050 a_0$ occurs due to the non-analytic behavior of $k$ and $E$ at the classical turning points. Additionally, the dashed orange curve Fig. \ref{fig:comparison} shows the lower-lying potential curve $E^{(1)}_\text{eff}$ based on the effective scattering length $a_\text{eff}[0]=-18.5 a_0$ which was found out to agree better with the experimentally observed spectra \cite{bendkowsky_observation_2009}.
As we will outline in Sec. \ref{sec:regular}, one reason for the discrepancy between $a_\text{eff}[0]$ and $a[0]$, which has not been fully understood so far, is the effect of couplings between non-degenerate $\ket{\varphi_i}$ and $\ket{\varphi_j}$ basis states that, at the level of first order perturbation theory, are neglected in (\ref{eqn:first_order_fermi}) but lead to deeper binding energies. This coupling becomes especially important in cases where the quantum defect of the state in question is close to an integer value.

To take the couplings into account, which means to analyze the system beyond the first order perturbation theory, we diagonalize $\hat{H}$ numerically in an extended basis set.
Technically this is done by fixing $k$ as in (\ref{eqn:k_fix}) to some energy $E_i^0$ and diagonalizing $\hat{H}$ in a basis set of states range-bound to a certain number of quantum defect eigenstates $\ket{n l m}$ whose energies lie close to $E_i^0$. In the case of $Rb(35s)+Rb(5s)$, we choose $E_i^0=E^0_{35,0,0}$. Furthermore, it is convenient to group the basis states into manifolds of similar energy, that we call $n^*$-manifolds. Each consists of the three quantum defect states $\ket{n^*+3,0,0}$,  $\ket{n^*+2,1,0}$, $\ket{n^*+1,2,0}$ and the $n^*-3$ energetically degenerate states $\ket{n^*,l>2,0}$. For practical purposes and without loss of generality, we can neglect states with $m\neq0$. The projection of the electronic angular momentum onto the internuclear $\vec{R}$ axis parallel to the $z$-axis is a good quantum number. States with $m\neq0$ have vanishing density on the $z$-axis and hence show no interaction via the potential $V$.

The resulting BO potential energy curves are shown in Fig. \ref{fig:comparison}: Using only the $n^*=32$ manifold gives the BO energy $E_0$ (cyan dots), including the $n^*$-manifolds with $|n^*-32|\leq 1$ leads to the BO energy $E_1$ (light blue dots) and taking $|n^*-32|\leq 5$ results in the BO energy $E_5$ (dark blue dashed-dotted line). 
Here every increase in the size of the basis set leads to considerably lower potential curves. Surprisingly, even for large basis sets consisting of up to 10 $n^*$-manifolds, the potential curves do not converge.

The following calculation shows that this non-convergence is not a numerical artifact but inherent to the bare delta function potential used for the electron-perturber interaction and the diagonalization procedure. We will illuminate this in the following. Let us consider the finite subspace $\mathcal{W}$ spanned by $N$ eigenstates $\ket{\varphi_i}$ and investigate the diagonalization of $\hat{H}$ inside $\mathcal{W}$ in more detail.

The projection of the Schr\"odinger equation
\begin{equation}
\hat{H} \Psi(\vec{r})= E \Psi(\vec{r})
\label{eqn:schroedinger}
\end{equation}
onto $\mathcal{W}$ reads
\begin{equation}
\sum_{j=1}^N \left[E^0_i \delta_{ij}+2 \pi a[k]\varphi_i^*(\vec{R})\varphi_j(\vec{R}) \right]\alpha_j = E \alpha_i 
\label{eqn:matrix}
\end{equation}
where we made the ansatz
\begin{equation}
\Psi(\vec{r})= \sum^N_{j=1} \alpha_j \varphi_j(\vec{r}) \ .
\label{eqn:psi_expansion}
\end{equation}
Consequently, (\ref{eqn:psi_expansion}) will lead to the exact solution of (\ref{eqn:schroedinger}) in the limit $N \to \infty$.

From  (\ref{eqn:matrix}) we deduce 
\begin{equation}
\alpha_i = \left(2 \pi a[k]  \sum_{j=1}^N \alpha_j \varphi_j(\vec{R}) \right)\frac{\varphi_i^*(\vec{R})}{E-E_i^0} = \mathcal{N} \frac{\varphi_i^*(\vec{R})}{E_i^0-E}
\end{equation}
where $\mathcal{N}$ is a normalization constant. Reinserting this into (\ref{eqn:matrix}) yields the characteristic equation
\begin{equation}
1+ 2 \pi a[k] \sum_{i=1}^N \frac{|\varphi_i(\vec{R})|^2}{E^0_i-E}=0 \ .
\label{eqn:energy_bare}
\end{equation}
The roots $E$ of (\ref{eqn:energy_bare}) are now exactly the eigenvalues that one would obtain by a numerical diagonalization of (\ref{eqn:matrix}).
However, in contrast to numerical diagonalization methods there is no need to perform the approximation (\ref{eqn:k_fix}) for the involved $k$-vector in numerical root finding algorithms. Hence, (\ref{eqn:energy_bare}) is even more general.

To see the effects of diagonalization inside large basis sets, we take the limit $N \to \infty$. 
The sum in (\ref{eqn:energy_bare}) can then be replaced by the Green's function of $\hat{H}^0$
\begin{equation}
G^0(\vec{r},\vec{r}',E)= \sumint{i=1}{\infty} \frac{\varphi^*_i(\vec{r}')\varphi_i(\vec{r})}{E_i^0-E} 
\label{eqn:G_expansion}
\end{equation}
evaluated in the limit $\vec{r}=\vec{r}'=\vec{R}$, where the sum above includes discrete and continuum states.
However, for $\vec{r} \to \vec{r}'$ the Green's function diverges as
\begin{equation}
G^0(\vec{r},\vec{r}',E) \propto \frac{1}{2 \pi |\vec{r} - \vec{r}'|} \ .
\label{eqn:G_divergence}
\end{equation}
Therefore the characteristic equation (\ref{eqn:energy_bare}) possesses for $N \to \infty$ no well defined solution $E$. This explains the nonconvergent behavior that we encounter with increasing size of the basis set for the diagonalization, as illustrated in Fig. \ref{fig:comparison}.
\section{Regularized delta function potential}
\label{sec:regular}
In the literature it is well known that the usefulness of a delta function potential in three dimensions is restricted since its action on irregular wave functions is not well defined and its spectrum is unbounded from below \cite{dalibard_collisional_1999,geltman_bound_2011}.
We employ therefore the regularized delta function potential
\begin{equation}
V(\vec{r},\vec{R})= 2 \pi a[k(R,E)] \delta(\vec{r}-\vec{R}) \frac{\partial}{\partial  \rho }  \rho ,
\label{eqn:delta_regular}
\end{equation}
whose solutions fulfill the condition (\ref{eqn:psi_semiclassic}) exactly \cite{huang_quantum-mechanical_1957}.
Using again the basis functions $\varphi_i$ we obtain
\begin{equation}
0 =(E^0_i-E) \alpha_i + 2 \pi a[k(R,E)] \varphi_i^*(\vec{R}) \left[ \frac{\partial}{\partial  \rho }\left(  \rho \sumint{j=1}{\infty} \alpha_j  \varphi_j(\vec{r})\right)\right]_{\vec{r} \to \vec{R}}
\label{eqn:matrixreg}
\end{equation}
Similar to (\ref{eqn:energy_bare}) and as shown in \cite{busch_two_1998} we derive that a nontrivial solution of (\ref{eqn:matrixreg}) exists only for energies satisfying 
\begin{equation}
1+ 2 \pi a[k(R,E)] \left[\frac{\partial}{\partial \rho} \left(\rho \sumint {i=0}{\infty} \frac{\varphi^*_i(\vec{R})\varphi_i(\vec{r})}{E^{0}_i-E}\right) \right]_{\vec{r} \to \vec{R}}=0 \ .
\label{eqn:energy_regularized}
\end{equation}
In contrast to (\ref{eqn:energy_bare}) this equation is not restricted to finite regions of the Hilbert space and yet well defined. It is crucial to carry out the summation before taking the derivative in (\ref{eqn:matrixreg}) and (\ref{eqn:energy_regularized}) as the opposite, i.e. shifting the derivative into the sum, would again lead to irregularities, due to (\ref{eqn:G_divergence}). To this end, we insert explicitly the Green's function (\ref{eqn:G_expansion}) into (\ref{eqn:energy_regularized})  which yields
\begin{equation}
1+2 \pi a[k(R,E)] G^{0}_\text{reg}(\vec{R},\vec{R},E)=0 \ ,
\label{eqn:energy_Green}
\end{equation}
where
\begin{equation}
G^{0}_\text{reg}(\vec{R},\vec{R},E):= \left[ \frac{\partial}{\partial \rho} \left( \rho G^{0}(\vec{r},\vec{R},E)\right)\right]_{\substack{\vec{r} \to \vec{R}}} \ .
\label{eqn:regularization} 
\end{equation}
This is exactly the result obtained alternatively via the quantum defect theory in \cite{khuskivadze_adiabatic_2002,hamilton_photoionization_2003} and agrees nicely with the expectation that the pseudopotential approach and the quantum defect approach should result in equivalent potential curves. Eq. (\ref{eqn:energy_Green}) can be evaluated numerically by expressing the Green's function in (\ref{eqn:G_expansion}) in terms of Whittaker functions, which is possible for arbitrary quantum defects \cite{davydkin_quadractic_1971}.
The BO potential energy curve  $E_\text{reg}(R)$ obtained in this manner for the $Rb(35s)+Rb(5s)$ state with the parameters (\ref{eqn:parameters}) is shown (orange dashed curve) in Fig \ref{fig:comparison}. Compared to the first order perturbation theory results, the outer minimum is around 30\%  deeper and shifted towards smaller $R$. This is an indication of the contribution of higher $l$-states and explains why the observed binding energies in \cite{bendkowsky_observation_2009} were larger than expected from the first order perturbation approximation (\ref{eqn:first_order_fermi}). In \cite{bendkowsky_observation_2009}, this was compensated partially by the introduction of an effective scattering length $a_\text{eff}[0]$.     

More analytical insight is gained by performing first order perturbation theory with (\ref{eqn:delta_regular}).
Additional care is necessary when using the regularized delta function potential in standard perturbation theory, due to the non-commutativity of summation and differential operations \cite{huang_quantum-mechanical_1957}. Nevertheless, the first order corrections $E_i^{(1)}$ to the energy $E_i^{0}$ can be obtained in the normal manner as
\begin{eqnarray}
E_i^{(1)}&=& 2 \pi a[k(R,E_i^{0}] \int d^3r \varphi^*_i(\vec{r})  \delta(\vec{r}-\vec{R}) \frac{\partial}{\partial  \rho }  \rho \varphi_i(\vec{r}) \nonumber \\
&=& 2 \pi a[k(R,E_i^{0}] \left|\varphi_i(\vec{R})\right|^2 
\label{eqn:firstorder}
\end{eqnarray}
where energetically degenerate $\varphi_i$ are chosen such that they diagonalize $V$ in the degenerate manifold. 
Hence the first order energies of the regularized delta function potential (\ref{eqn:firstorder}) coincide with the first order energies of the bare delta function potential (\ref{eqn:first_order_fermi}). Therefore, even though we showed that the bare delta function potential is in general not adequate, it serves as a valid first order approximation for the regularized delta function potential.

More generally, Fig. \ref{fig:comparison} suggests that the bare delta function potential may even approximate the regularized delta function potential beyond the first order perturbation theory, by including only a few more, but not too many, adjacent states. For example the BO potential energy curve $E_0$ obtained by including only states adjacent to the $n=32$ manifold, lies close to $E_\text{reg}(R)$.
Here a more rigorous comparison between the gain {in} accuracy and the error accumulated when increasing the size of the basis set, would be desirable and should be carried out in the future.      
This would be especially useful as the regularized delta function potential can not be implemented in a numerical diagonalization procedure that relies on the basis functions $\varphi_i$ which are regular at $R$. We see this from (\ref{eqn:matrixreg}), which, in contrast to (\ref{eqn:matrix}), can not be converted into a system of linear equations for the coefficients $\alpha_i$ because the sum has to be carried out before taking the differential.

\section{Summary \& Conclusion}
Aspects of the numerical implementation of the Fermi pseudopotential approach, relevant to the calculation of the BO potential energy curves of the ultralong-range Rydberg molecules bound by pure $s$-wave scattering are delineated. It is shown that the bare delta pseudopotential leads to unphysical results that diverge with increasing basis set size in any diagonalization scheme.
This behavior is described analytically. The convergent BO potential curves, that agree exactly with the results obtained within the quantum defect theory can be produced by employing a regularized delta function potential. Although the bare delta function potential still yields the correct binding energies in the first order perturbation theory, the example of the $Rb(35s)+Rb(5s)$ molecule indicates that beyond the first order effects are not negligible. By including some of the neighboring Rydberg basis states, some of these corrections can be reproduced.  However it remains to be seen how many additional states can rigorously be included in exact diagonalization schemes and whether the zero-range pseudopotential for the $p$-wave-interaction \cite{hamilton_shape-resonance-induced_2002}, will require similar regularization as well.

\ack
The authors thank C. H. Greene and W. Li for fruitful discussions. C.F. gratefully acknowledges a scholarship by the Studienstiftung
des deutschen Volkes.

%%%%%%%%%%%%%%%%%%%%%%%%%%%%%%%%%%%%%%%%%%%%%%%%%%%%%%%%%%%%%%%%%%%%%%%%%%%%%%%%

%\bibliography{biblio}
%\bibliographystyle{science}
%\bibliographystyle{apsrev4-1}
\end{document}